\documentstyle[preprint,prd,aps]{revtex}

\def\lsim{\begin{array}{c} < \\ \sim \end{array}}
\begin{document}
\draft
\title{\large \bf A small but nonzero cosmological constant \thanks{This
essay received an honorable mention in the Annual Essay Competition of the 
Gravity Research Foundation for the year 2000}} 
\author{\bf Y. Jack Ng$^{(a),(b)}$
and H. van Dam$^{(b)}$}
\address{(a) Center for Theoretical Physics, Laboratory for Nuclear
Science and Department of Physics,\\
Massachusetts Institute of
Technology,
Cambridge, MA 02139\\}
\address{(b) Institute of Field Physics, Department of Physics
and Astronomy,\\
University of North Carolina, Chapel Hill, NC 27599-3255\\}
\maketitle
\bigskip

\begin{abstract}

Recent astrophysical observations seem to indicate that the cosmological
constant is small but nonzero and positive.  The old cosmological
constant problem asks why it is so small; we must now ask, in addition, 
why it is nonzero (and is in the range found by recent observations), and 
why it is positive.  In this essay, we try to
kill these three metaphorical birds with one stone.  That stone is the
unimodular theory of gravity, which is the ordinary theory of gravity,
except for the way the cosmological constant arises in the theory.  We
argue that the cosmological constant becomes dynamical, and eventually, in
terms of the cosmic scale factor $R(t)$, it takes the form $\Lambda(t) =
\Lambda(t_0) (R(t_0)/R(t))^2$, but not before the epoch corresponding to
the redshift parameter $z \sim 1$.

\end{abstract}

\newpage

Until recent years, there used to be only one well-known
problem\cite{weing} with the
cosmological constant, viz., why it is so small --- some 120 orders of
magnitude smaller than what we naively think it should be.  If it is that
small, it must be zero, so some of us thought.  Now we know better.  The
recent astrophysical observations indicate that, quite likely, the
cosmological constant is not zero, though small, and positive, giving rise
to cosmic repulsion.\cite{quint}  We must now
ask these additional questions:
Why is the cosmological constant not zero?  Why does it have the observed
magnitude, contributing to the energy density of the observable universe 
about twice as much as matter?

In this essay, we will attempt to present a qualitative solution to
these three problems of the cosmological constant.  We will do so in the
framework of unimodular gravity\cite{bij,nvd,ein,ht,sor97}
 which, as we will show, is nothing but the
ordinary theory of gravity --- except for one curious twist which has to
do with the way the cosmological constant arises in the theory.

First let us reiterate the cosmological constant problems and put them in
a form that will be useful later in the essay.  From the Einstein-Hilbert 
action of gravity, we know that the cosmological constant $\Lambda$ has
units of the reciprocal of length squared.  Until recent years, all
galactic observations had failed to detect any spacetime distortions that
one can attribute to a nonzero cosmological constant out to the
farthest distance, about $10^{28}$ cm., in the observable universe.
Denote the 4-volume of the observable universe by $V$, then the empirical
observations give the bound $\Lambda \lsim V^{-1/2}$.  But theoretical
expectations would predict a much larger value: $\Lambda \sim l_P^{-2}$
with $l_P \sim 10^{-33}$cm being the Planck length.  This vast discrepancy
by 122 orders of magnitude constitutes the old cosmological constant
problem: why is $\Lambda$ so small?  Recent observations\cite{bops} 
(supernovae 1a, cosmic microwave background, cluster density and evolution
etc) are consistent with a geometrically flat universe and they indicate
that the cosmological constant contributes about 70\% of the energy
density; hence
\begin{equation}
\Lambda \sim  + \frac{1}{\sqrt{V}},
\label{obs}
\end{equation}
the cosmological constant is non-zero (and positive) after all.

Two observations are now in order.  First, it is not surprising that
$\Lambda$ is non-zero since setting $\Lambda = 0$ does not enhance the
existing symmetry of the gravitation theory.  Second, the (old)
cosmological constant problem is insensitive to the non-renormalizability
of quantized general relativity as the problem occurs well below the
Planck scale (even the relatively small vacuum energy density in QCD
yields a discrepancy of about 42 orders of magnitude).  Thus it seems
reasonable that one can adequately address the cosmological constant
problems in the framework of a gravity theory whose classical limit
resembles general relativity.  In the following, we consider the
unimodular theory of gravity.

Unimodular gravity is actually very well motivated on physical grounds. 
Following Wigner\cite{wigner} for a proper quantum description of the
massless spin-two graviton, the mediator in gravitational
interactions, we naturally arrive at the concept of gauge
transformations.\cite{bij}
Without loss of generality, we can choose the graviton's two polarization 
tensors to be traceless (and symmetric).  But since the trace of the
polarization states is preserved by all the transformations, it is natural
to demand that the graviton states be described by \emph{traceless}
symmetric tensor fields.  The strong field generalization of the traceless
tensor field is a metric tensor $g_{\mu \nu}$ that has unit determinant:
$-det g_{\mu \nu} \equiv g = 1$, hence the name "unimodular gravity."

At first sight, the unimodular constraint has greatly changed the
gravitational field equation,
since now only the traceless combinations appear:
\begin{equation}
R^{\mu \nu} - 1/4 g^{\mu \nu} R = -8 \pi G (T^{\mu \nu} - 1/4 g^{\mu \nu}
T^{\lambda}_{\lambda}),
\label{less}
\end{equation}
where $T^{\mu \nu}$ is the conserved matter stress tensor.  But in
conjunction with the Bianchi identity for the covariant derivative of the
Einstein tensor, the field equation yields 
$D^{\mu}(R - 8 \pi G T^{\lambda}_{\lambda}) = 0$.  Thus $(R - 8 \pi G
T^{\lambda}_{\lambda})$ is a constant.  Denoting that constant
of integration by $-4 \Lambda$, we find
\begin{equation}
R^{\mu \nu} - 1/2 g^{\mu \nu} R = \Lambda g^{\mu \nu} - 8 \pi G T^{\mu
\nu},
\label{eins}
\end{equation}
the familiar Einstein's equation.  The only difference from the ordinary
theory is in the way $\Lambda$ arises in the theory --- it is an
(arbitrary)
integration constant, unrelated to any parameter in the original action.
There are two other differences\cite{bij} that are worth mentioning.
(1) Unlike the
ordinary theory, the Lagrangian for unimodular gravity can be expressed
as a polynomial of the metric field.  (2) Conformal transformations
$g_{\mu
\nu} = C^{2} g'_{\mu \nu}$ in the unimodular theory of gravity are very
simple, the unimodular constraint fixes the conformal factor $C$ to 
be 1.   

Since $\Lambda$ arises as an arbitrary constant of integration, it has
\emph{no} preferred value classically.  In the corresponding quantum
theory, we expect the state vector of the universe to be given by a
superposition of states with different values of $\Lambda$ and the
quantum vacuum functional to receive contributions from all
different values of $\Lambda$.  For the quantum theory, it is advantageous
to start with a generalized version of the
classical unimodular theory given above, that is generally covariant while
preserving
locality.  We will use the version of unimodular gravity given by the 
Henneaux and Teitelboim action\cite{ht}
\begin{equation}
S_{unimod} = - \frac{1}{16 \pi G} \int [ \sqrt{g} (R + 2 \Lambda) - 2
\Lambda  
\partial_\mu {\mathcal T}^\mu](d^3x)dt.
\label{HT}
\end{equation}
One of its equations of motion is $\sqrt{g} = \partial_\mu
\mathcal{T}^\mu$,
the generalized unimodular condition, with $g$ given in terms of the 
auxiliary field $\mathcal{T}^{\mu}$.  Note that, in this theory, 
$\Lambda$ plays the role of
"momentum" conjugate to the "coordinate" $\int d^3x {\mathcal T}_0$ which
can
be identified, with the aid of the generalized unimodular condition, as
the spacetime volume $V$\cite{Vconst}.  Hence $\Lambda$ and $V$ are
conjugate to each other.

We are ready to argue why the observed cosmological constant is so small.
The argument\cite{nvd} makes crucial use of quantum mechanics.  Consider
the vacuum
functional for unimodular gravity given by path integrations over
$\mathcal{T}^{\mu}$, $g_{\mu \nu}$, the matter fields (represented
by $\phi$), and $\Lambda$:
\begin{equation}
Z_{Minkowski} = \int d\mu (\Lambda) \int d [\phi] d [g_{\mu \nu}] \int d
[{\mathcal T}^{\mu}]  exp \left\{ -i[ S_{unimod} + S_{M}(\phi, g_{\mu
\nu})]\right\}, 
\label{Z}
\end{equation}
where $S_{M}$ stands for the contribution from matter fields (and $d \mu
(\Lambda)$ denotes the measure of the $\Lambda$ integration).  The
integration over $\mathcal{T}^{\mu}$ yields $\delta(\partial_{\mu}
\Lambda)$, which implies that $\Lambda$ is spacetime-independent (befiting
its role as the cosmological constant).  A Wick rotation now allows us to
study the Euclidean vacuum functional $Z$.  The integrations over $g_{\mu
\nu}$ and $\phi$ give $exp[-S_{\Lambda}(\overline{g}_{\mu \nu},
\overline{\phi})]$ where $\overline{g}_{\mu \nu}$ and $\overline{\phi}$
are the background fields which minimize the effective action
$S_{\Lambda}$.  A curvature expansion for $S_{\Lambda}$ yields a
Lagrangian whose first two terms are the Einstein-Hilbert terms $\sqrt{g}
(R + 2\Lambda)$, where $\Lambda$ now denotes the \emph{fully renormalized}
cosmological constant.  We can make a change of variable from the original
(bare) $\Lambda$ to the renormalized $\Lambda$.  Let us assume that for
the present cosmic era, $\phi$ is essentially in the ground state, then
it is reasonable to neglect the effects of $\overline{\phi}$.  To
continue, we follow Baum\cite{baum} and Hawking\cite{hawking} to evaluate
$S_{\Lambda}(\overline{g}_{\mu \nu}, 0)$.  For negative $\Lambda$,
$S_{\Lambda}$ is positive; for positive $\Lambda$, one finds
$S_{\Lambda}(\overline{g}_{\mu \nu}, 0) = -3 \pi / G\Lambda$, so that 
\begin{equation}
Z = \int \! d\mu (\Lambda) exp(3 \pi /G \Lambda).
\label{finalZ}
\end{equation}
The essential singularity of the integrand at $\Lambda = 0 +$ means that
the overwhelmingly most probable configuration is the one with $\Lambda =
0$, and this in turn implies that the observed cosmological constant in
the present era is essentially zero.

There is one serious shortcoming in the above argument involving the Wick 
rotation to Euclidean space.  It is well-known
that the Euclidean formulation of quantum gravity is plagued by the
conformal factor problem, due to divergent path-integrals.  In our
defense, we want to point out that we have used the effective action in
the Euclidean formulation at its stationary point only.  We should
also recall that 
the conformal factor problem is arguably rather benign in the
original version of unimodular gravity (as pointed out above), so perhaps
it is not that serious even in the generalized version that we
have just employed.  There is another cause for concern.  Since part of
the above argument bears some
resemblance to Coleman's wormhole approach\cite{coleman}, one may worry
that some of the objections to Coleman's argument (on top of the conformal
factor problem) may also apply here.  Fortunately, it appears that they
do not.\cite{avoid}  In any case, to the extent that our argument is
valid, we have understood why the observed cosmological constant is so
small and why, if the cosmological constant is not exactly zero, it is
positive.

In the above argument, we have assumed that for the present cosmic era,
the matter fields are in their ground states so that their effects on the
effective action can be neglected; and the end result is that the
observed $\Lambda$ is zero.  Plausible as this assumption is, it is
not entirely correct.  So, we do expect a non-vanishing (but
small) cosmological
constant for the present era, and $\Lambda$ goes to zero only
asymptotically as the universe expands.  Regrettably, we have not been able
to calculate the small but not-entirely-negligible effects of the matter
fields on the effective action.  We will adopt the attitude that the above
result is valid to the lowest order of approximation for which $\Lambda$
is zero.  We will now borrow an argument due to Sorkin\cite{sor97} to make
an order of magnitude estimate of the cosmological constant (to the
next leading order).\cite{qualify}

There are two ingredients to Sorkin's argument.  First, from unimodular
gravity he takes the idea that $\Lambda$ is in some sense conjugate to the
spacetime volume $V$.  Hence their fluctuations obey a Heisenberg-type
quantum uncertainty principle,
\begin{equation}
\delta V \! \delta \Lambda \sim 1,
\label{heisenb}
\end{equation}
where we have used the natural units ($\hbar = 1$, $G = 1$).  The second
ingredient to Sorkin's argument does not seem to be directly related to the
unimodular theory
of gravity.  It is drawn from the causal-set theory\cite{reid}, which
stipulates that continous geometries in classical gravity should be
replaced by "causal-sets", the discrete substratum of spacetime.  The
fluctuation in the number of elements $N$ making up the set is of the
Poisson type, i.e., $\delta N \sim \sqrt{N}$.  For a causal set, the
spacetime volume $V$ becomes $N$.  It follows that
\begin{equation}
\delta V \sim \delta N \sim \sqrt{N} \sim \sqrt{V}.
\label{poisson}
\end{equation}
Putting Eqs. (\ref{heisenb}) and (\ref{poisson}) together yields a minimum
uncertainty in 
$\Lambda$\ of $\delta \Lambda \sim V^{-1/2}$.\cite{why}  But we have 
already argued that $\Lambda$ vanishes to the lowest order of
approximation and that it is positive if it is not zero.  So we conclude
that $\Lambda$ fluctuates about zero with a magnitude of $V^{-1/2}$ and it
is positive:
\begin{equation}
\Lambda \! \sim \! + \frac{1}{\sqrt{V}},
\label{wow}  
\end{equation}
which, lo and behold, is Eq. (\ref{obs})!  The cosmological
constant is small, but non-zero and positive, and has the correct order of
magnitude as observed.  In other words, $\Lambda$ contributes to the
energy of the
universe an amount on the order of the critical density.  As a side remark,
we note that if we now appeal to
the inflationary universe scenario, we may also understand why matter
contributes a comparable amount.\cite{coincid} 
  
We emphasize that $\Lambda$ is of the form given by Eq. (\ref{wow}) only
after the matter fields have, more or less, settled down to the ground
state.  To be more precise about the epoch when $\Lambda$ starts taking on
that form, we consider the Friedmann-Robertson-Walker cosmologies.  In that
case, Eq. (\ref{wow}) becomes\cite{wei}
\begin{equation}
\Lambda(t) = \Lambda(t_0) \left(\frac{R(t_0)}{R(t)}\right)^{2}.
\label{wow2}
\end{equation}
For the flat case which our universe appears to approximate, the expansion
rate in the post-radiation-dominated era is given by \cite{CDFN}
\begin{equation}
\frac{1}{H_0^2} \left(\frac{\dot{R}}{R}\right)^2 = (1 - 3
{\Omega}_{\Lambda})
w^3 + 3 {\Omega}_{\Lambda} w^2,
\label{FRW}
\end{equation}
where ${\Omega}_{\Lambda}$ is the fractional energy density in the present
era due to the
cosmological constant (Eq. (\ref{wow2})), $ 0 < w \equiv R_0 / R = 1 + z <
\infty$,
and $H_0$ is
the current Hubble parameter.  A cosmic bounce occurs whenever the
right-hand side has a zero for a real positive $w$.  The root is given by
$w = 3 {\Omega}_{\Lambda} / (3 {\Omega}_{\Lambda} - 1)$, which, for the
observed
value\cite{bops} of ${\Omega}_{\Lambda} = 7/10$, equals 21/11.  It
follows that the absence of a
bounce in the past restricts the allowed redshift parameter to $z \leq
10/11$.  We conclude that the cosmological constant is of the form
given by Eq. (\ref{wow2}) only after the cosmic epoch corresponding to $z
\lsim 1$ (the formation of clusters of galaxies).  Eventually $\Lambda$
dominates the cosmic evolution, driving the
universe to expand at the rate given by $R \sim t$.

In summary, we have proposed that one can understand, in the framework of
the unimodular theory of gravity, why the cosmological constant
is so small but non-zero and positive, and is in the range found by recent 
astrophysical observations.
This theory is well motivated, its original form being 
based on the quantum
description of helicity-two particles.  It leads to a theory of
gravity in which the cosmological constant is freed to become dynamical 
at the quantum level, and its form is given by Eq. (\ref{wow2}) for $z
\lsim 1$.  To work out the form of the dynamical cosmological constant for
the earler epochs will be the next challenge.

\bigskip

This essay is based mainly on the talk given by one of us (YJN) in the 1999
DESY
Theory Workshop on ``$\nu$s from the Universe.''  We thank H. Nielsen for 
encouraging us to write it up.
YJN thanks W. Buchmuller for the 
hospitality extended to him at DESY.  He also thanks W. Buchmuller, N. Dragon, 
H. Nielsen, and especially R. D. Sorkin and P. H. Frampton for useful
discussions.  
The work was supported in part by
the U.S. Department of Energy under \#DF-FC02-94ER40818 and
\#DE-FG05-85ER-40219,
and by the Bahnson Fund of the University of North Carolina at Chapel Hill.
YJN was on leave at MIT where this essay was written.  
He thanks the faculty at the Center
for Theoretical Physics for their hospitality.


\bigskip
\begin{references}

\bibitem{weing}
For a review of the problem, see S. Weinberg, Rev. Mod. Phys. {\bf 61}, 1
(1989), Y. J. Ng, Int. J. Mod. Phys. {\bf D1}, 145 (1992), and S. M. 
Carroll, W. H. Press, and E. L. Turner, Ann. Rev. Astron. \& Astrophys.
{\bf 30}, 499 (1992).  Note that, very likely, there was a large early 
cosmological constant driving cosmic inflation.  But here we want to explain 
why the effective cosmological constant is small \emph{now}, not why it was 
always small.

\bibitem{quint}
We will not discuss other ideas and mechanisms (like quintessence) that 
claim to explain the astrophysical data. 

\bibitem{bij}
J. J. van der Bij, H. van Dam, and Y. J. Ng, Physica {\bf A116}, 307
(1982).

\bibitem{nvd}
Y. J. Ng and H. van Dam, Phys. Rev. Lett. {\bf 65}, 1972 (1990).

\bibitem{ein}
A. Einstein, in The Principle of Relativity, ed. A. Sommerfeld (Dover, New
York, 1952); J. L. Anderson and D. Finkelstein, Am. J. Phys. {\bf 39}, 901
(1971); J. Rayski, Gen. Relativ. Gravit. {\bf 11}, 19 (1979); S. Weinberg,
unpublished (1983); F. Wilczek, Phys. Rep. {\bf 104}, 111 (1984); A. Zee,
in Proceedings of the Twentieth Annual Orbis Scientiae on High Energy
Physics, 1985, eds. S. L. Mintz and A. Perlmutter (Plenum, New York,
1985); W. Buchmuller and N. Dragon, Phys. Lett. {\bf B207}, 292 (1988); W.
G. Unruh and R. M. Wald, Phys. Rev. {\bf 40}, 2598 (1989); J. D. Brown and
J. W. York, Jr., Phys. Rev. {\bf D40}, 3312 (1989); R. D. Sorkin, Int. J. 
Th. Phys. {\bf 33}, 523 (1994); A. N. Petrov,
Mod. Phys. Lett. {\bf A6}, 2107 (1991); Izawa K.-I., Prog. Theor. Phys.
{\bf 93}, 615 (1995).

\bibitem{ht}
M. Henneaux and C. Teitelboim, Phys. Lett. {\bf B222}, 195 (1989).

\bibitem{sor97}
R. D. Sorkin, in Relativity and
Gravitation: Classical and Quantum, eds. J. C. D'Olivo et al. (World
Scientific, Singapore, 1991); Int. J. Th. Phys. {\bf 36}, 2759 (1997).

\bibitem{bops}
For a review of the recent data, see N. A. Bahcall, J. P. Ostriker, S.
Perlmutter, and P. J. Steinhardt, Science {\bf 284}, 1481 (1999), and 
references therein.

\bibitem{wigner}
E. P. Wigner, Ann. Math. {\bf 40}, 149 (1939).

\bibitem{Vconst}
We have taken the additive constant term to be $V = 0$ at $t = 0$. 

\bibitem{baum}
E. Baum, Phys. Lett. {\bf B133}, 185 (1984).

\bibitem{hawking}
S. W. Hawking, Phys. Lett. {\bf B134}, 403 (1984).

\bibitem{coleman}
S. Coleman, Nucl. Phys. {\bf B310}, 643 (1988).

\bibitem{avoid}
Since we have a simple (instead of Coleman's double) exponential in the
integrand for the vacuum functional in Eq. (\ref{finalZ}) in the text,
Polchinski's argument (J. Polchinski, Phys. Lett. {\bf B219}, 251 (1989))
against the wormhole approach is not applicable here.  Also, Duff's
argument (M. Duff, Phys. Lett. {\bf B226}, 36 (1989)) against the
Baum-Hawking approach is not applicable here because our argument does
\emph{not} involve the step that Duff (correctly) finds objectionable. 

\bibitem{qualify}
Sorkin's formulation and interpretation of the unimodular theory differs
somewhat from ours.  See Sorkin in Ref.\cite{ein}.  Here we merely make use of 
some of his ideas.

\bibitem{reid}
For an introduction to the causal-set theory, see, e.g., Ref.\cite{sor97}.

\bibitem{why}
Note that the fluctuations in the renormalized $\Lambda$ are given entirely 
by those in the bare $\Lambda$, since only the bare $\Lambda$, as the 
conjugate to $V$, undergoes this kind of quantum fluctuations.  This result 
is consistent with Sorkin's prediction (see Ref. \cite{sor97}).  On the other 
hand, we should point out that the dynamics of the causal-set theory is not 
yet fully understood; so the last part of the argument may not be on solid 
ground.

\bibitem{coincid}
Thus we may have ameliorated the ``cosmic coincidence'' problem of why both 
matter and the cosmological constant contribute comparable amounts to the 
energy density of the universe in the present era. 

\bibitem{wei}
$\Lambda$ of this form was proposed by W. Chen and Y. S. Wu, Phys. Rev.
{\bf D41}, 695 (1990), based on other reasons.

\bibitem{CDFN}
J. L. Crooks, J. O. Dunn, P. H. Frampton, and Y. J. Ng, astro-ph/0005406.

\end{references}
\end{document}